\documentclass[aps,nofootinbib,prd,superscriptaddress,eqsecnum,showpacs,showkeys,preprintnumbers,altaffilletter]{revtex4}
\usepackage{amsfonts,amsmath,amssymb,amsthm,bbm,hyperref}
\usepackage{graphicx}
\usepackage{color}
\usepackage[T1]{fontenc}
\usepackage[normalem]{ulem}

\def\D{\mathrm{d}}
\def\I{\mathrm{i}}
\def\E{\mathrm{e}}
\def\del{\partial}
\def\be{\begin{equation}}
\def\ee{\end{equation}}

\newcommand{\MP}{M_\mathrm{P}}

\newcommand{\beq}{\begin{eqnarray}}
\newcommand{\eeq}{\end{eqnarray}}
\newcommand{\nn}{\nonumber}

\newcommand{\cn}{\mathrm{cn}}
\newcommand{\sn}{\mathrm{sn}}

\newcommand{\Kmax}{K_\text{max}}
\newcommand{\HdS}{H_\mathrm{dS}}
\newcommand{\cHdS}{{\mathcal{H}}_{\mathrm{dS}}}

\newcommand{\kn}{\kappa}

\newcommand{\txi}{\tilde{\xi}}

\begin{document}

\title{What if? -- Exploring the Multiverse through Euclidean wormholes}

\author{Mariam Bouhmadi-L\'{o}pez}
\email{mariam.bouhmadi@ehu.eus}
\affiliation{Department of Theoretical Physics, University of the Basque Country UPV/EHU, P.O.~Box 644, 48080 Bilbao, Spain}
\affiliation{IKERBASQUE, Basque Foundation for Science, 48011 Bilbao, Spain}

\author{Manuel Kr{\"a}mer}
\email{manuel.kraemer@usz.edu.pl}
\affiliation{Institute of Physics, University of Szczecin, Wielkopolska 15, 70-451 Szczecin, Poland}

\author{Jo\~ao Morais}
\email{jviegas001@ikasle.ehu.eus}
\affiliation{Department of Theoretical Physics, University of the Basque Country UPV/EHU, P.O.~Box 644, 48080 Bilbao, Spain}

\author{Salvador Robles-P\'{e}rez}
\email{salvarp@imaff.cfmac.csic.es}
\affiliation{Instituto de  F\'{\i}sica Fundamental, CSIC, Serrano 121, 28006 Madrid, Spain}
\affiliation{Estaci\'{o}n Ecol\'{o}gica de Biocosmolog\'{\i}a, Pedro de Alvarado 14, 06411 Medell\'{\i}n, Spain}

\date{\today}

\begin{abstract}
We present Euclidean wormhole solutions describing {\it possible bridges} within the multiverse. The study is carried out in the framework of the third quantization. The matter content is modelled through a scalar field which supports the existence of a whole collection of universes. The instanton solutions describe Euclidean solutions that connect baby universes with asymptotically de Sitter universes. We compute the tunnelling probability of these processes. Considering the current bounds on the energy  scale of inflation and assuming that all the baby universes are nucleated with the same probability, we draw some conclusions about what are the universes more likely to tunnel and therefore undergo a standard inflationary era.
\end{abstract}


\maketitle

%
%

\section{Introduction}

Humankind has, ever since history can tell us, been looking for possible answers and hints to the questions: (i) where do we come from? and (ii) where are we heading to? Cosmology is the path to address these questions on a scientific ground. In what refers to the first question, general relativity predicts the existence of a past big bang, at least for standard matter and for a homogeneous and isotropic universe, a singularity which is hoped to be wiped out  through a primordial quantum era \cite{clausbook}. At the semiclassical level, this may imply the existence of Euclidean solutions or instantons that geometrically describe Euclidean wormholes or bridges in the spacetime, on its widest sense, where the big bang singularity is circumvented or at least shadowed by the presence of Euclidean wormholes \cite{Hawking:1987mz,Giddings:1987cg,Halliwell:1989ky,Barcelo:1995gz,mariam2002,RP2014} connecting baby universes to some Lorentzian singularity-free universes. 

While Euclidean wormholes can be seen as a natural geometrical and gravitational extension of the Coleman--De Luccia instanton \cite{Coleman:1980aw}, where the latter assumes the existence of two vacua and a decaying process connecting them, the former do not necessarily assume the existence of such vacua. In both cases, however, an extremization of the Euclidean action is reached. In addition, Euclidean wormholes are not exclusive to general relativity and can be found, for example, in string theory (see, for example, \cite{ArkaniHamed:2007js,Hertog:2017owm}). Moreover, even though in general one uses the term ``Euclidean wormhole'' to describe a curved Euclidean space with -- at least -- two large asymptotic regions connected by a throat, this idea has been extended as well to spacetimes with a cosmological constant that can be positive \cite{mariam2002,chen16} or negative \cite{Barcelo:1995gz,barcelo97,barcelo98}. We would like to highlight that strictly speaking, a Euclidean wormhole -- in the simplest geometrical setup -- is constructed from two glued instantons \cite{barvinsky06,barvinsky07}. However, given that in both cases Euclidean wormholes and instantons are Euclidean bridges connecting different Lorentzian spacetimes, we will use the words ``instanton'' and ``wormhole'' indistinguishably.

It is therefore natural to assume the existence of wormholes as {\textit{connecting paths}} not only within remote regions of the universe but also within the multiverse as a whole \cite{PFGD2011}. In this regard, we would like to stress that the paradigm of inflation, supported observationally since COBE's first measurements \cite{Mather:1993ij} of the Cosmic Microwave Background anisotropies and the subsequent experiments WMAP \cite{Hinshaw:2012aka} and Planck \cite{Ade:2015xua}, predicts and supports the existence of the multiverse. In fact, the multiverse concept can be reached and understood from several approaches, from the seminal idea of Everett of a multiverse formed by the branches of quantum mechanics \cite{Everett1957}, to the landscape of the string theories \cite{Susskind2003}, the inflationary multiverse \cite{Linde1983, Linde1986}, or the ekpyrotic scenario \cite{Khoury2001, Steinhard2002}, among many others \cite{Carr2007, Smolin1997, Tegmark2003, Freivogel2004, Mersini2008b, Mersini2008a, mariam2007, RP2007b, Alonso2012}. In any case, one assumes the existence of an undetermined number of realizations of the universe, each one causally separated from the others by the presence of quantum barriers, event horizons or extra dimensions. Yet, their quantum states may still be related by the existence of non-local correlations in the global quantum state of the spacetime and the matter fields.

In this paper, we show the existence of wormhole solutions in the framework of the third quantization, one of the current proposals to describe the multiverse. It basically consists of considering the solution of the Wheeler--DeWitt equation as a field that propagates in the minisuperspace of spacetime metrics and matter fields, and thus quantizing the wave function of the spacetime and matter fields by following a formal parallelism with the customary procedure of a quantum field theory (see Refs.~\cite{Caderni1984, McGuian1988, Strominger1990, RP2010}). Then, the creation and annihilation operators of the third quantization formalism describe the creation (or the annihilation) of a particular spacetime-matter  configuration. In particular, the solutions that we have found correspond to {\it{Euclidean tunnels}} that connect baby universes with asymptotically de Sitter universes. In this regard, we have focussed exclusively on the tunnelling between two Lorentzian universes supported by the same scalar field with the same initial kinetic energy. A more complete analysis would consider the ``communication'' or tunnelling between universes with different initial kinetic energies for the scalar field. A further possibility to describe the ``communication'' between different universes of the multiverse is by considering entangled universes \cite{cyclic,salva2017}.

The paper can be outlined as follows. In section II, we summarise the model we will be analysing and review briefly the third-quantization approach.  Then, we find for the first time exact solutions describing an instanton within this framework in presence of a minimally coupled massive scalar field. In this framework and within a semiclassical approach, the multiverse can be seen as a collection of semiclassical universes with a label that indicates the initial kinetic energy of the scalar field that supports them. In section III, for a given universe within the multiverse; i.\,e.~a given label, we obtain the transition probability describing  the tunnelling from a baby universe to an asymptotically de Sitter universe. Our calculations assume the tunnelling boundary conditions of Vilenkin \cite{Vilenkin:1984wp, Vilenkin}. Finally, in section IV, we present our conclusions. For clarity, we include as well an appendix where we obtain analytically the transition amplitude analysed in section III.

%
%

\section{Model}

We consider a closed FLRW universe with scale factor $a$ containing a minimally coupled scalar field $\varphi$ with mass $m$ and quadratic potential $\mathcal{V}(\varphi) = \frac{1}{2}\,m^2\varphi^2$. The Wheeler--DeWitt (WDW) equation for the wave function $\phi(a,\varphi)$ of such a configuration of the spacetime and matter field\footnote{The wave function $\phi(a,\varphi)$ represents the quantum state of the spacetime and the matter fields all together. Hence, the traditional name \emph{wave function of the universe} \cite{Hartle1983}. However, this name can be misleading in the multiverse scenario we are working on, so we shall just call it the wave function of the spacetime and matter fields.} reads \cite{Linde:2005ht}
\be \label{basicwdw}
\Biggl[\frac{\hbar^2 G}{3\pi}\,\frac{\del^2}{\del a^2} - \frac{\hbar^2}{4\pi^2 a^2}\,\frac{\del^2}{\del \varphi^2} - \frac{3\pi a^2}{4 G}+2a^4\pi^2 \,\mathcal{V}(\varphi) \Biggr]\phi(a,\varphi) = 0\,.
\ee
Here, we have used a specific choice of factor ordering. Choosing an alternative ordering would introduce an additional term with a first derivative of  $\phi$ with respect to $a$, but such a term would not influence our calculations.

We can simplify this equation and absorb several constants by rescaling the scalar field as 
\be
\varphi \rightarrow \sqrt{\frac{4\pi G}{3}} \,\varphi\,,
\ee
such that $\varphi$ becomes dimensionless. If we furthermore define the quantity
\be
\omega^2(a,\varphi) := \sigma^2 \left(H_\varphi^2 a^4 - a^2\right),
\ee
which contains the definitions
\be
\label{sigma_def}
H_\varphi^2 := \frac{8 \pi G}{3}\,\mathcal{V}(\varphi) \quad \text{and} \quad \sigma := \frac{3 \pi}{2 G}\,,
\ee
we end up with the following simpler form of the WDW equation \eqref{basicwdw}:
\be
\label{parentwdw}
\hbar^2\,\frac{\del^2\phi}{\del a^2} - \frac{\hbar^2}{a^2} \frac{\del^2 \phi}{\del \varphi^2} + \omega^2(a,\varphi) \phi = 0\,.
\ee
In order to model the multiverse, we use the third quantization formalism, which essentially consists of promoting the wave function of the spacetime and matter fields, $\phi(a,\varphi), $ to an operator, $\hat{\phi}(a,\varphi) $, given in the case being considered by 
\be\label{modedecomp}
\hat{\phi}(a,\varphi) = \int \frac{\D K}{\sqrt{2\pi}} 
\left[\E^{\I K \varphi} \phi_K(a) \, \hat{b}_{K} + \E^{-\I K \varphi} \phi_K^*(a) \, \hat{c}^\dag_{K}\right],
\ee
where $\hat{b}_{K}$ and $\hat{c}^\dag_{K}$ are the annihilation and creation operators, respectively, of universes whose evolution will be specified later on. The modes $K$ are related to the momentum conjugated to the scalar field, $p_\varphi$, and we interpret the decomposition (\ref{modedecomp}) in the way that each amplitude $\phi_K(a)$ of the  wave function $\phi(a,\varphi)$ represents a single universe with a specific value of $p_\varphi$. The wave functions of the universes satisfy then the following effective WDW equation
\be \label{subwdw}
\hbar^2\,\frac{\del^2\phi_K}{\del a^2} + \omega_K^2 \phi_K = 0\,,
\ee
where $\omega_K$ is given by
\be \label{omk}
\omega_K(a) := \sigma \sqrt{a^4 \HdS^2 - a^2 + \frac{\hbar^2 K^2}{\sigma^2 a^2}}\,.
\ee
Here, $\HdS$ is a constant that arises from specifying $H_\varphi$ to a specific value of $\varphi$. We can also see that in the effective WDW equation \eqref{subwdw} describing the individual universes, the $\varphi$-derivative term, $- \frac{\hbar^2}{a^2} \frac{\del^2 \phi}{\del \varphi^2}$, appearing in \eqref{parentwdw} was converted into $\frac{\hbar^2 K^2}{a^2}$. In addition, it can be shown that the evolution of the universes follows the effective Friedmann equation \cite{salva2011,Garay:2013pba}
\be \label{modFri}
H^2 \equiv \left(\frac{\dot a}{a}\right)^2 = \frac{\omega^2_K(a)}{\sigma^2a^4} = \HdS^2 - \frac{1}{a^2} + \frac{\hbar^2 K^2}{\sigma^2 a^6}\,.
\ee
This behaviour with the additional term $\propto a^{-6}$ can be related to the model of an interacting multiverse described in \cite{salva2015}. However, we have not introduced here any explicit interaction between the universes and the last term in (\ref{modFri}) appears solely from the consideration of the quantum character of the mode decomposition (\ref{modedecomp}). It is thus a pure quantum correction term {without classical analogue%
\footnote{A stiff matter content of the universe would introduce a similar term, proportional to $a^{-6}$, in the Friedmann equation. However, the term here is also proportional to $\hbar$, which reveals its quantum nature without classical analogue ($\hbar\rightarrow 0$).}.}

In order to illustrate the evolution of this universe, we write down the radicand in \eqref{omk} in terms of its roots $a_+ \geq a_- \geq a_0$,
\be\label{OmgK01}
\omega_K(a) = \frac{\sigma \HdS}{a} \sqrt{(a^2 - a_+^2)(a^2 - a_-^2)(a^2 + a_0^2)}\,,
\ee
where 
\beq 
a_+\left(K\right) &:=& \frac{1}{\sqrt{3}\HdS} \sqrt{1 + 2 \cos\!\left(\frac{\alpha_K}{3}\right)} \,, \label{root+} \\
a_-\left(K\right) &:=& \frac{1}{\sqrt{3 } \HdS} \sqrt{1 - 2 \cos\!\left(\frac{\alpha_K + \pi}{3}\right)} \,, \label{root-}\\
a_0\left(K\right) &:=& \frac{1}{\sqrt{3}\HdS} \sqrt{-1 + 2 \cos\!\left(\frac{\alpha_K - \pi}{3}\right)} \,, \label{root0}
\eeq
and
\be\label{alpha}
\alpha_K 
:= {\rm arccos}\!\left(1 - 2\frac{K^2}{K_\text{max}^2}\right) 
{= 2 \arcsin\left(\frac{ K}{K_\text{max}}\right)}
\in [0,\pi] \,.
\ee\\
The maximum value of $K$ appearing in the last expression is defined as
\be\label{kmax}
	K_\text{max} 
	:= 
     \frac{\pi}{\sqrt{3}}\,\frac{M_\text{P}^2}{\hbar^2\,\HdS^2}
	= \frac{\pi}{\sqrt{3}}\frac{1}{\gamma}
	\,,
\ee
where we have introduced the Planck mass $M_\text{P}^2 := \hbar/G$ and the ratio $\gamma:=\hbar^2\HdS^2/\MP^2$ that relates the scale of inflation to the Planck mass.

\begin{figure}
  \centering
\includegraphics[width=0.5\linewidth]{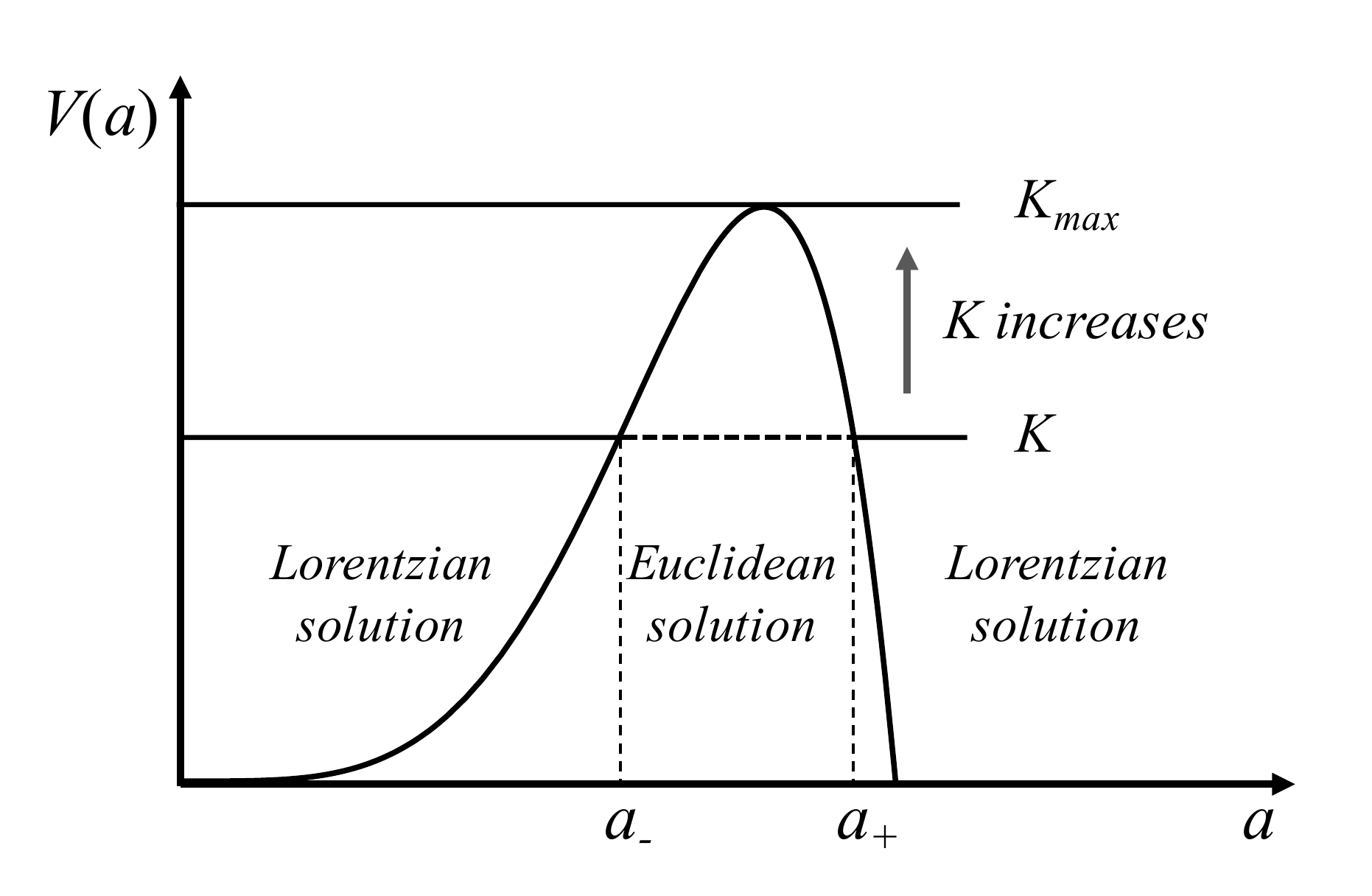}
\caption{The tunnelling potential $V(a) = \sigma^2\left(a^4-\HdS^2a^6\right)$.}
\label{potfig}
\end{figure}

From (\ref{modFri}) and (\ref{OmgK01}), we obtain the picture of a universe that behaves as a recollapsing baby universe for $a < a_-$, and as an asymptotically de Sitter universe for $a>a_+$. In between, $a_-<a<a_+$, there is a Euclidean, classically forbidden, region. This is similar in spirit to the model discussed in \cite{mariam2002} and can be illustrated by plotting the potential
\be
V(a) =\sigma^2\left(a^4-\HdS^2a^6\right),
\ee
which is depicted in Fig.~\ref{potfig}.

In order to obtain the evolution $a(\eta)$ of the different phases of the universe in terms of the conformal time $\eta$, defined in terms of the cosmic time $t$ via $\D\eta/\D t = a^{-1}$, we need to solve the  following differential equation
\be
\frac{\D a(\eta)}{\D \eta} 
= \frac{\omega_K(a)}{\sigma }\,,
\ee
where we have chosen the expanding branch and which translates into integrating
\be
\D\eta = \frac{{ \sigma}\D a}{\omega_K(a)}\,. 
\ee
We redefine the variable to be integrated over and use the definitions
\be \label{defx}
x := a^2, \quad x_+ := a^2_+, \quad x_-:= a^2_-, \quad x_0 := a^2_0\,,
\ee
such that we obtain
\be \label{d_eta_final}
\D\eta = \frac{\D x}{2 \HdS \sqrt{(x - x_+)(x - x_-)(x + x_0)}}\,.
\ee
A solution for the previous equation in terms of elementary functions can be obtained for the special cases of $\alpha_K=0$, which corresponds to the scenario of the creation of an expanding universe from nothing \cite{Vilenkin:1984wp}, and of $\alpha_K=\pi$, which corresponds to the maximum value of $K$ for which the tunnelling effect happens. In the remaining part of this section, we present the solutions of Eq.~\eqref{d_eta_final} in the Lorentzian regions $0<x<x_-$ and $x_+<x$ and in the Euclidean region $x_-<x<x_+$ for the non-trivial cases $\alpha_K\in(0,\,\pi)$.

%
%

\subsubsection{Baby Universe: $0 < x < x_-$}

The behaviour of the baby universe in the Lorentzian region $0 < x < x_-$ can be obtained by employing the change of variable\footnote{We point out that this transformation is not valid for $\alpha_K=0$, since in that case $x_0=x_-=0$ and the argument inside the $\arccos$ on the right-hand side of Eq.~\eqref{xminus_transformation} diverges. Physically this can be understood by the fact that for $\alpha_K=0$ there is no baby universe -- the expanding asymptotically de Sitter universe is created from nothing \cite{Vilenkin:1984wp}.} (cf.~Eq.~17.4.63 in Ref.~\cite{abra}):
\begin{align}
	\label{xminus_transformation}
	x \rightarrow\xi_- 
	:=&~ \arccos\left(\sqrt{\frac{(x_+-x_-)(x+x_0)}{(x_-+x_0)(x_+-x)}}\right),
\end{align}
with $\xi_-$ decreasing monotonically with $x$ from a maximum value $\xi_-(x=0)= \arccos[({\sin[(\pi-\alpha_K)/6]/\cos[\alpha_K/6]})^{1/2}]$ to $\xi_-(x=x_-)=0$.
Upon substitution in Eq.~\eqref{d_eta_final} and after an integration from $\eta$ to $\eta_-:=\eta(x_-)$, we find that the conformal time fulfills
\begin{align}
	\label{BabyUniverse_eta}
	\cHdS\left(\eta_- - \eta\right)
	=
	F\left(\xi_- \,\middle|\, k^ 2 \right),
\end{align}
where $F(\xi | m )$ is the elliptical integral of the first kind as defined in \cite{abra} and where we have introduced the notation
\begin{align}
	\label{cHdS_and_k_def}
	 {\cHdS} 
	 := \sqrt{x_0+x_+}\,\HdS
	\,
	 \qquad
	 \textrm{and}
	 \qquad
	k  
	:=\sqrt{\frac{x_- + x_0}{x_+ + x_0}}
	\,.
\end{align}
In order to obtain the evolution of the scale factor as a function of the conformal time we can use the relation of the elliptic integrals with the Jacobi elliptic functions $\mathrm{cn}(u|m)$ and $\mathrm{sn}(u|m)$ \cite{abra} to invert Eq.~\eqref{BabyUniverse_eta}. After some algebra we obtain:
\begin{align}
	a^2(\eta) 
	=&~
	{
	a_-^2 
	 - (a_0^2 + a_-^2)\,
	 \frac{\sn^2\left[{\cHdS}\left(\eta_--\eta\right) \,\middle|\,k^ 2\right]}
	 {1 - k^2\,\sn^2\left[{\cHdS}\left(\eta_--\eta\right) \,\middle|\, k^ 2\right]}
	 }
	\,.
\end{align}

%
%

\subsubsection{Asymptotically de Sitter Universe: $x_+ < x < +\infty$}

The evolution of the approximate de Sitter universe in the Lorentzian region $x_+ < x < +\infty$ can analogously be derived by employing the change of variable  (cf.~Eq.~17.4.62 in Ref.~\cite{abra}):
\begin{align}
	x \rightarrow\xi_+ 
	:=&~ \arcsin\left(\sqrt{\frac{x-x_+}{x-x_-}}\right),
\end{align}
with $\xi_+$ growing from $\xi_+(x=x_+)=0$ to  $\xi_+(x\rightarrow+\infty)=\pi/2$.
Upon substitution in Eq.~\eqref{d_eta_final} and after integrating from $\eta_+=\eta(x_+)$ to $\eta$ we find that the conformal time fulfills:
\begin{align}
	\label{deSitterUniverse_eta}
	\cHdS\left(\eta - \eta_+\right)
	 =&~
	 F\left(\xi_+ \,\middle|\, k^ 2\right).
\end{align}
The solution \eqref{deSitterUniverse_eta} can be inverted to obtain the scale factor as a function of the conformal time
\begin{align}
	a^2(\eta) 
	=&~ a_+^2 + (a_+^2-a_-^2)
	\frac{\sn^2\!\left[{\cHdS} \left(\eta - \eta_+\right)\,\middle|\, k^2\right]}
	{\cn^2\!\left[{\cHdS} \left(\eta - \eta_+\right)\,\middle|\, k^ 2\right]}
	\,.
\end{align}

%
%

\subsubsection{Euclidean Wormhole: $x_- < x < x_+$}

In order to obtain the solution for the Euclidean wormhole in the region $x_- < x < x_+$, we use the change of variable\footnote{We point out that this transformation is not valid for $\alpha_K=\pi$, since in that case $x_+=x_-$ and the argument inside the $\arccos$ on the right-hand side of Eq.~\eqref{Wormhole_xi} diverges. Physically this can be understood by the fact that for $\alpha_K=\pi$ there is no Euclidean region.} (cf.~Eq.~17.4.69 in Ref.~\cite{abra}):
\be
	\label{Wormhole_xi}
	x\rightarrow\txi := \arccos\left(\sqrt{\frac{x-x_-}{x_+-x_-}}\right)
	\,.
\ee
The new variable $\txi $ decreases monotonically with $x$ from $\txi(x=x_-)=\pi/2$ to $\txi(x=x_+)=0$. When replacing \eqref{Wormhole_xi} in \eqref{d_eta_final} and integrating in Euclidean time $\tilde\eta=\I\eta$ from $\tilde\eta$ to $\tilde\eta_+=\tilde\eta(x=x_+)$ we find that the conformal Euclidean time fulfills
\begin{align}
	\label{Wormhole_eta}
	\cHdS\left(\tilde\eta_+ - \tilde\eta\right)
	 =&~
	F\left(\txi \,\middle|\,  1-k^ 2\right)
	 \,.
\end{align}
By inverting the solution \eqref{Wormhole_eta}, we obtain the expression for the scale factor as a function of the conformal time:
\begin{align}
	a^2(\eta) 
	= &~
	a_+^2 - \left(a_+^2 - a_-^2\right)\mathrm{sn}^2\!\left[{\cHdS} \left(\tilde\eta_+ - \tilde\eta\right)\,\middle|\, 1-k^2 \right]
		\,.
\end{align}
Please notice that our solution generalises the Giddings--Strominger instanton \cite{Giddings:1987cg}, even though the two solutions have a completely different origin (for further applications of this instanton, see e.g.~\cite{sengupta14}). In our solution, constructed in the framework of the third quantization, the appearance of a term in the Friedmann equation that scales as $a^{-6}$ is due to the quantization scheme, while in the Giddings--Strominger instanton, which is supported by an axion whose field strength tensor is defined through a rank-three anti-symmetric tensor $H_{\mu\nu\lambda}$, a similar term appears in the Friedmann equation by the fact that $H_{\mu\nu\lambda}$, which contributes quadratically to the action, is subjected to the constraint $\D H =0$ \cite{Giddings:1987cg}.

\begin{figure}
  \centering
\includegraphics[width=.8\linewidth]{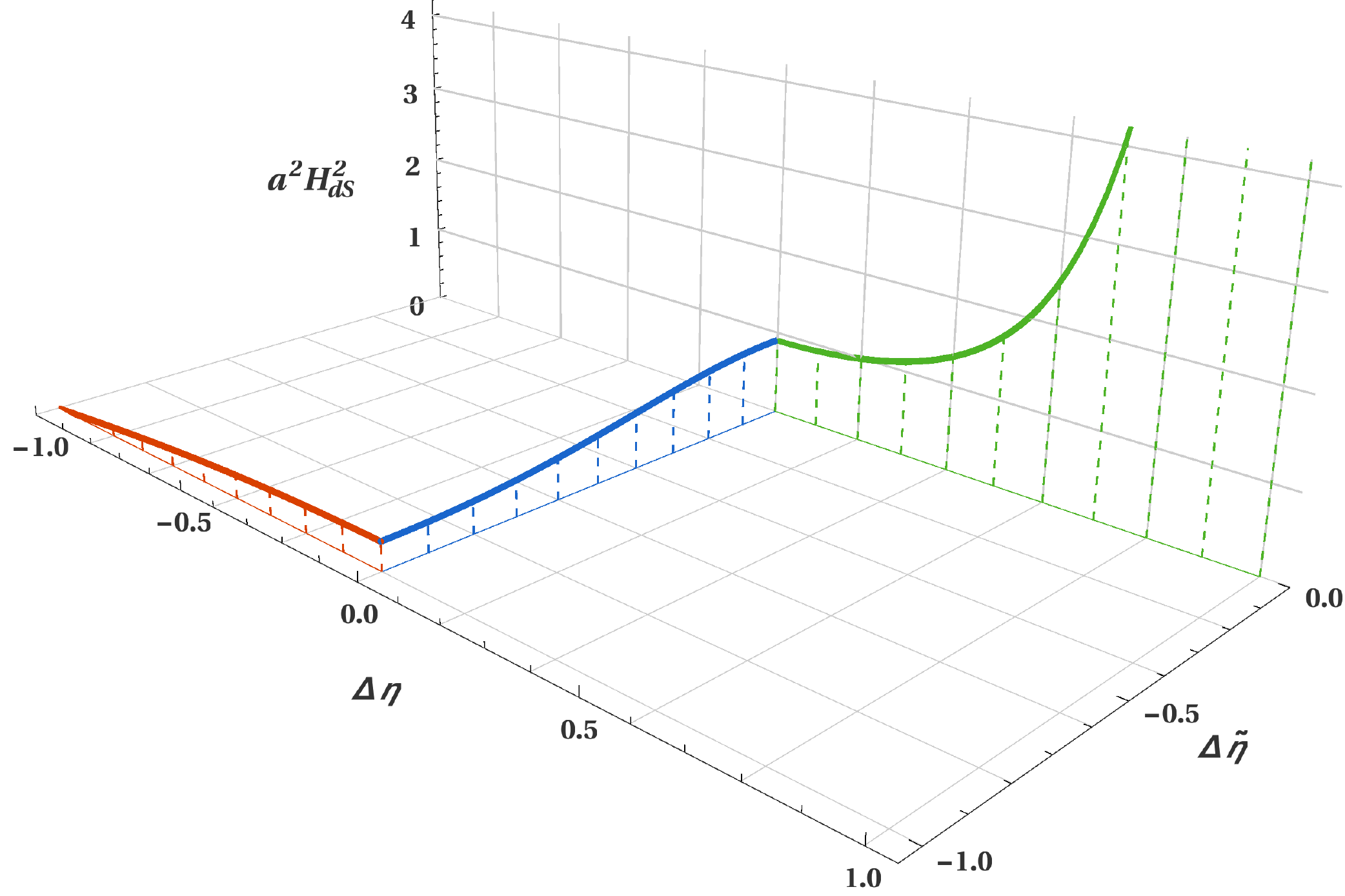}
\caption{\label{a2_evolution}The evolution of the squared scale factor as a function of the conformal Lorentzian time $\eta$ and conformal Euclidean time $\tilde\eta$.
In order to be able to plot the evolution of the scale factor in a single figure including the Lorentzian and Euclidean solutions, we have rescaled the conformal time as follows: for the baby universe (red) $\Delta\eta = (\eta - \eta_-)/|\eta_{(a=0)} - \eta_-|$ with $\Delta\eta\in[-1,0]$; for the Euclidean instanton (blue) $\Delta\tilde\eta = (\tilde\eta - \tilde\eta_+)/|\tilde\eta_{(a=a_-)} - \tilde\eta_+|$ with $\Delta\tilde\eta\in[-1,0]$; for the expanding asymptotically de Sitter universe (green) $\Delta\eta = (\eta - \eta_+)/|\eta_{(a\rightarrow+\infty)} - \eta_+|$ with $\Delta\eta\in[0,1]$.
}
\end{figure}

{ In Fig.~\ref{a2_evolution} we depict the combined evolution of the squared scale factor during the two Lorentzian regions and through the Euclidean wormhole.
During the baby universe phase (depicted in red), the scale factor evolves from $0$ to $a_-$ as the time displacement $\Delta\eta:=(\eta-\eta_-)/|\eta_{(a=0)}-\eta_-|$ varies from $-1$ to $0$.
As the scale factor reaches the value $a_-$, the universe can enter a Euclidean wormhole (depicted in blue), in which the scale factor grows from $a_-$ to the maximum value $a_+$ as the Euclidean time displacement $\Delta\tilde\eta:=(\tilde\eta - \tilde\eta_+)/|\tilde\eta_{(a=a_-)}-\tilde\eta_+|$ goes from $-1$ to $0$.
Once the value $a_+$ is reached, the universe exits the Euclidean wormhole and enters a near de Sitter expansion (depicted in green). In this final phase the scale factor grows in an accelerated fashion as the time displacement $\Delta\eta:=(\eta-\eta_+)/|\eta_{(a=+\infty)}-\eta_+|$ varies from $0$ to $1$.
}

%
%

\section{Tunnelling}

We now want to calculate the probability that the universe can quantum-mechanically tunnel from the baby universe phase to the de Sitter phase assuming the tunnelling transition as proposed in \cite{Vilenkin:1984wp,Vilenkin}: 
\be \label{probk}
	\mathcal{P}_K(a_-\rightarrow a_+) \approx \exp\left(-\,\frac{2\sigma \HdS}{\hbar} \int_{a_-}^{a_+}\D a\,\left|\frac{1}{a} \sqrt{(a^2 - a_+^2)(a^2 - a_-^2)(a^2 + a_0^2)}\right|\right).
\ee
Therefore, in order to calculate the tunnelling probability with respect to the value of $K$, we need to evaluate the following integral
\begin{align}
	 \label{Int_01}
	I :=&~  \int_{x_-}^{x_+} \frac{\D x}{2x}\,\sqrt{(x_+ - x) (x-x_-) (x+x_0) } 
	\,,
\end{align}
where $x$, $x_+$, $x_-$ and $x_0$ are defined in \eqref{defx}.
The integral \eqref{Int_01} becomes trivial for the special case of $K=0$, the creation of an expanding universe from nothing, as computed in \cite{Vilenkin:1984wp}. In this case we find $I_{(K=0)} = (1/3)x_+^{3/2}$.

For a general $K\in(0,\,\Kmax)$, the integral $I$ can be solved by means of the transformation 
\begin{align}
	x\rightarrow y := \sqrt{\frac{x_+ - x}{x_+ - x_-}}\,.
\end{align}
After some lengthy algebra, we find that we can express $I$ as a linear combination of the complete elliptic integrals of the first, second and third kind \cite{abra}, $K(m)$, $E(m)$, $\Pi(n|m)$, respectively:
\begin{align}
	 \label{tunnelint}
	I 
	=&~  \left(x_+ + x_0\right)^{3/2}
	\left[
		 C_K\, K\left(\tilde{k}^2\right) 
		 + C_E\, E\left(\tilde{k}^2\right) 
		 + C_\Pi \, \Pi\left(\kappa^2\,\middle|\,\tilde{k}^2\right)
	\right],
\end{align}
where  we have introduced the parameters
\begin{align}
	\tilde{k}:=\sqrt{\frac{x_+ - x_-}{x_+ + x_0}}
	\qquad
	\textrm{and}
	\qquad
	\kn := \sqrt{\frac{x_+ - x_-}{x_+}}
	\,.
\end{align}
The linear coefficients $C_K$, $C_E$ and $C_\Pi$ are defined as
\begin{align}
	C_K := \tilde{k}^2\left[
		\frac{1}{3}+\frac{1}{\kappa^2} 
		+ \tilde{k}^2\left(\frac{1}{3}-\frac{1}{\kappa^4}\right)
	\right]
	\,,
	\quad
	C_E := - \tilde{k}^2\left[
		\frac{1}{3}
		+ \tilde{k}^2\left(\frac{1}{3} - \frac{1}{\kappa^2} \right)
	\right]
	\,,
	\quad
	C_\Pi := \tilde{k}^2\left( 1- \frac{\tilde{k}^2}{\kappa^2}\right)\left( 1- \frac{1}{\kappa^2}\right)
	\,.
\end{align}
For the benefit of the reader, a detailed derivation of \eqref{tunnelint} is presented in the Appendix~\ref{App_A}, while a cross-check of the result using the differentiation of $I$ is presented in Appendix~\ref{App_crosscheck}.

\begin{figure} [t]
\begin{minipage}{.7\textwidth}
\includegraphics[width=\linewidth]{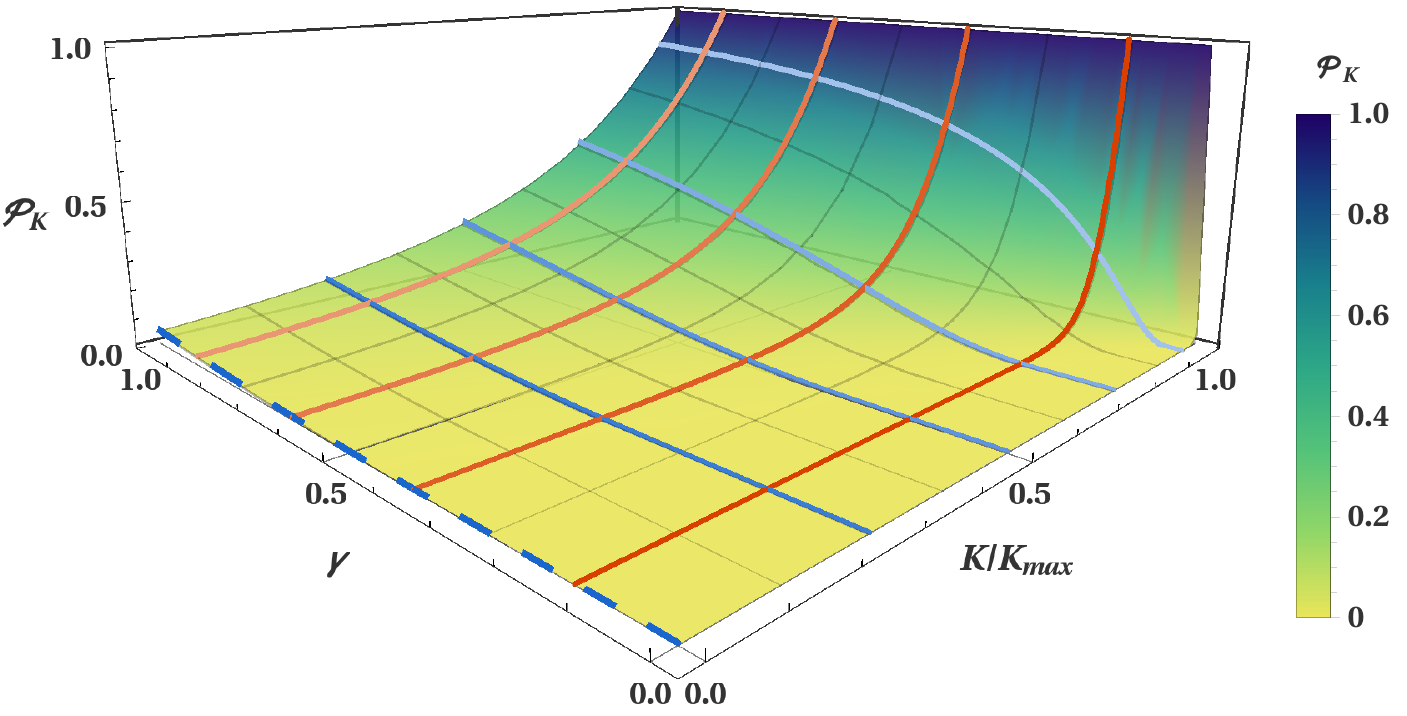}
\end{minipage}
\caption{\label{tunfig}The tunnelling probability $\mathcal{P}_K(a_-\rightarrow a_+)$ plotted as a function of the ratios $\gamma := \hbar^2\HdS^2/M_\text{P}^2$ and $K/K_\mathrm{max}$. The coloured lines, which represent the tunnelling probability for a fixed value of $K/\Kmax$ (blue) or of $\gamma$ (red) are compared in Fig.~\ref{tunfig2}. The tunnelling probability for the case of the creation of an expanding universe from nothing ($K=0$) is indicated by a dashed blue line.}
\vspace{10pt}
\begin{minipage}{.495\textwidth}
\includegraphics[width=\linewidth]{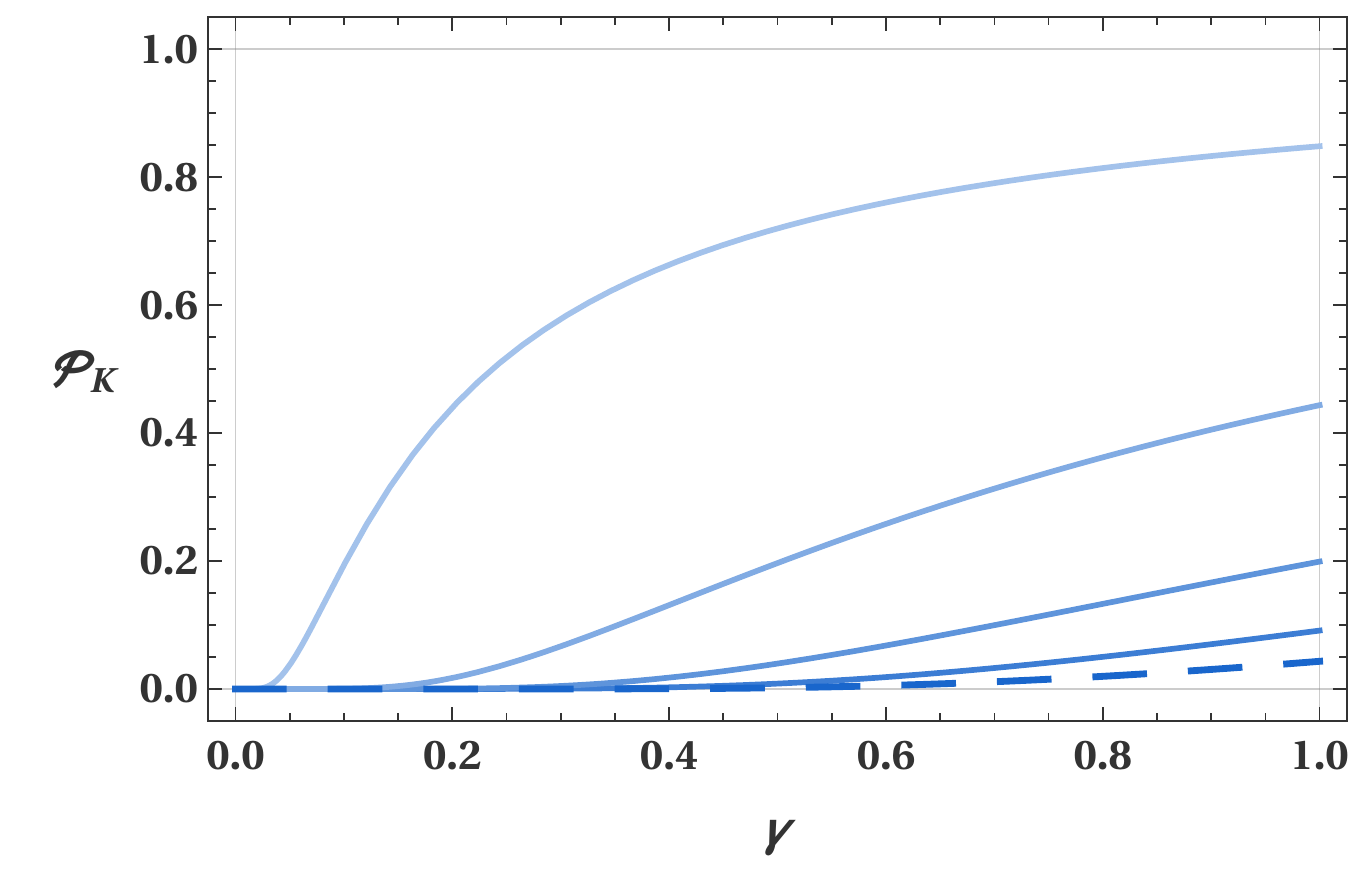}
\end{minipage}
\hfill
\begin{minipage}{.495\textwidth}
  \centering
\includegraphics[width=\linewidth]{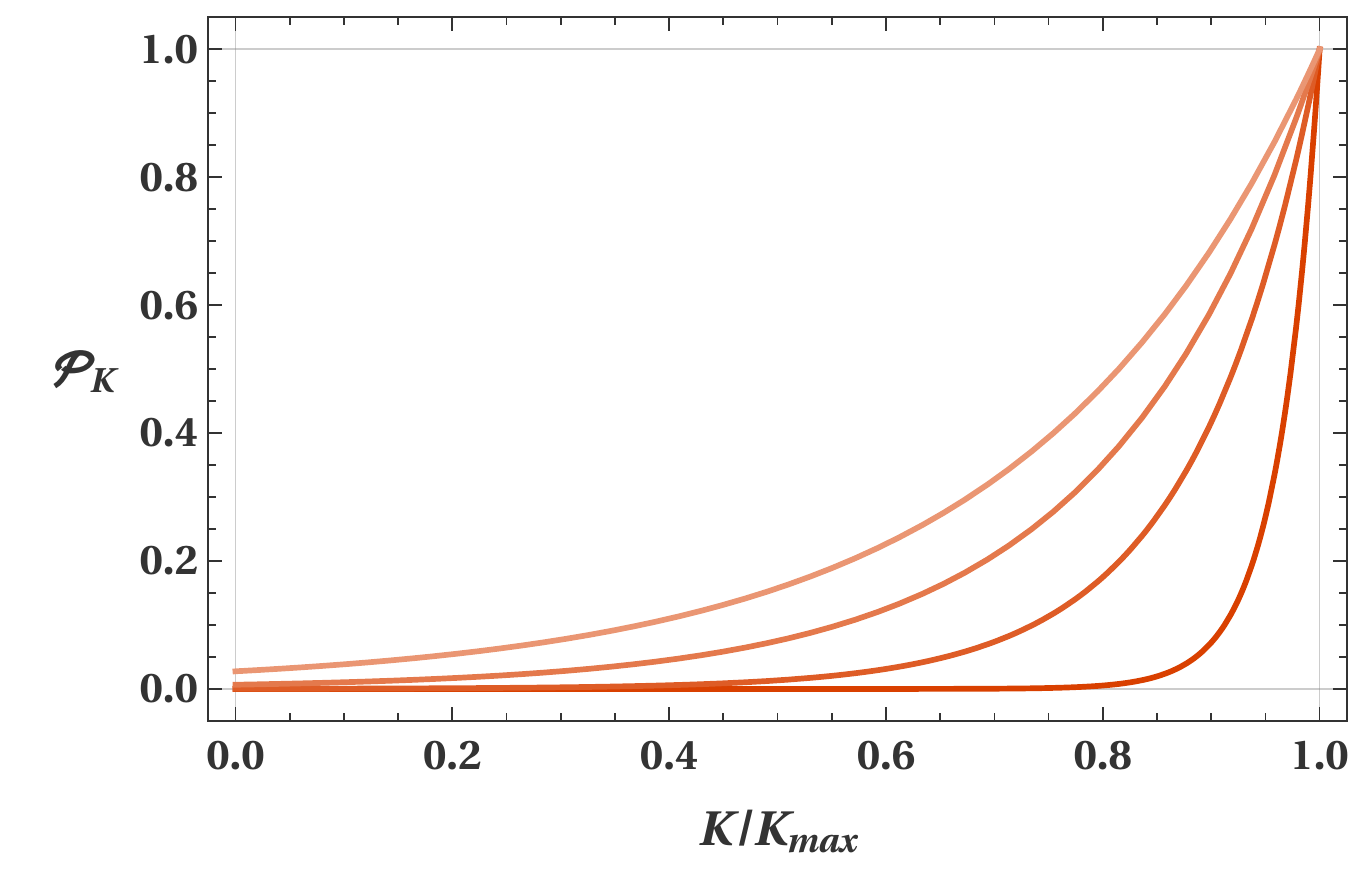}
\end{minipage}
\caption{\label{tunfig2}The tunnelling probability $\mathcal{P}_K(a_-\rightarrow a_+)$ plotted as (left)  a function of $\gamma$ for different values of $K/K_\mathrm{max}$: (from bottom/darker to top/lighter) $K/K_\mathrm{max}=0$, $K/K_\mathrm{max}=1/4$, $K/K_\mathrm{max}=1/2$, $K/K_\mathrm{max}=3/4$ and $K/K_\mathrm{max}=99/100$; and as (right) a function of the ratio $K/K_\mathrm{max}$ for different values of $\gamma := \hbar^2\HdS^2/M_\text{P}^2$: (from bottom/darker to top/lighter) $\gamma=1/8$, $\gamma=3/8$, $\gamma=5/8$ and $\gamma=7/8$. The tunnelling probability for the case of the creation of an expanding universe from nothing ($K=0$) is indicated by a dashed blue line.}
\end{figure}

Finally, after inserting \eqref{tunnelint} into \eqref{probk} and using the definitions \eqref{sigma_def} and \eqref{cHdS_and_k_def} to eliminate $\sigma$ and $\HdS$ in favour of the dimensionless parameters $\gamma$ and $\cHdS$ we can write the tunnelling probability as
\begin{align}
	\mathcal{P}_K(a_-\rightarrow a_+) 
	\approx
	\begin{cases}
	\exp\left(-\dfrac{\pi}{\gamma}\right)
	&
	\text{for } K=0
	\,,
	\\
	 \exp\left(
		-\dfrac{3\pi}{\gamma}\cHdS^3
		\left[
		 C_K\, K\left(\tilde{k}^2\right) 
		 + C_E\, E\left(\tilde{k}^2\right) 
		 + C_\Pi \, \Pi\left(\kappa^2\,\middle|\,\tilde{k}^2\right)
	\right]
	\right)
	&
	\text{for } 0<K<\Kmax
	\,.
	\end{cases}
\end{align}
Here, we note that the dimensionless parameters $\cHdS$, $\tilde{k}$ and $\kappa$, as well as the linear coefficients $C_K$, $C_E$ and $C_\Pi$, depend only on the ratio $K/\Kmax$ (through the angle $\alpha_K$). Consequently, $\mathcal{P}_K(a_-\rightarrow a_+)$ is a bivariable function of  $\gamma:=\hbar^2\HdS^2/\MP^2$, the ratio between the inflationary scale and the Planck scale, and of  $K/\Kmax$, i.\,e.~how close the scalar field momentum $K$ of the baby universe is to the maximum quantum allowed value. In Fig.~\ref{tunfig} we plot the tunnelling probability as a function of $\gamma$ and $K/\Kmax$, while in Fig.~\ref{tunfig2} we present the tunnelling probability  as a function of $\gamma$ for different ratios $K/\Kmax$ and as a function of $K/\Kmax$ for different values of $\gamma$. 
One can see that the tunnelling probability goes to 1 for $K \rightarrow \Kmax$, as expected, since the Euclidean region ceases to be present in that limit. For $K\approx0$ the tunnelling probability approaches the solution for the creation of an expanding universe from nothing \cite{Vilenkin:1984wp}, which is marked by a blue dashed line.
Due to the $1/\gamma$ factor in the argument of the exponential, the tunnelling probability decays rapidly for low values of $\gamma$. As such, if the scale of inflation is well below the Planck scale, then the tunnelling probability is extremely low except for $K\approx\Kmax$, as can be observed in Fig.~\ref{tunfig}.

%
%

\section{Conclusions}

Within the framework of the third quantization, one of the current proposals to describe the multiverse, we have shown the existence of Euclidean wormhole solutions which describe possible bridges within the multiverse. 

More precisely, by considering a massive  minimally coupled scalar field within the framework of the third quantization, we can describe a whole bunch of universes fingerprinted by the initial kinetic energy of the scalar field that supports them. It turns out that for a given initial kinetic energy of the scalar field two classically disconnected solutions emerge: a baby universe and an asymptotically de Sitter universe. Although these two Lorentzian solutions are classically disconnected, it turns out that this is no longer the case from a semi-classical point of view. In fact, as we have shown, these two solutions are connected through a Euclidean wormhole (cf.~Fig.~\ref{a2_evolution} for a schematic representation).
In addition, our solution generalises the Giddings--Strominger instanton \cite{Giddings:1987cg}, even though the two solutions have a completely different origin. While our solution is constructed in the framework of the third quantization with a massive minimally coupled scalar field,  the Giddings--Strominger instanton is constructed in the framework of string theory and is supported by an axion whose field strength tensor is defined through a rank-three anti-symmetric tensor.

Assuming the transition amplitude between two Lorentzian universes proposed in \cite{Vilenkin:1984wp,Vilenkin}, we have calculated the probability of tunnelling from the baby universe to the asymptotically de Sitter universe. Our results are graphically represented in Figs.~\ref{tunfig} and \ref{tunfig2}. What we can conclude is that the larger the initial kinetic energy of the scalar field is, the higher is the probability of the baby universe to cross the barrier depicted in Fig.~\ref{potfig}; i.\,e.~the higher is the probability that the universe crosses towards the inflationary era through the shortcut provided by the Euclidean wormhole.

Finally given that the highest value of the potential barrier (see Fig.~\ref{potfig}) separating the two Lorentzian universes is related to the scale of inflation, i.\,e.~$\HdS$ (cf.~for example Eq.~(\ref{omk})), of the asymptotically de Sitter universe, we can estimate $K_{\max}$ defined in Eq.~(\ref{kmax}) or equivalently the parameter $\gamma$ given in the same equation. In fact, given that the energy scale of inflation is at most of the order $8.8\times 10^{19}$ GeV  \cite{Bartolo:2016ami}, we can conclude that 
$\gamma$  must be smaller than $5.2029\times10^{-11}$ or equivalently  $K_{\max}$  must be quite large. Therefore, what we have proven is that if all the baby universes are nucleated with the same probability, those with larger $K$ are the most likely  to tunnel through the wormhole and therefore undergo an inflationary era  
like our own patch of the universe. In a subsequent paper, we will constrain the current model with CMB data following our previous works \cite{mariam2011,mariam2013}.

\section*{Acknowledgements}
This article is based upon work from COST Action CA15117 ``Cosmology and Astrophysics Network for Theoretical Advances and Training Actions (CANTATA)'', supported by COST (European Cooperation in Science and Technology). The work of M.\,B.-L.~is supported by the Basque Foundation of Science Ikerbasque. She and J.\,M.~also wish to acknowledge the partial support from the Basque government Grant No.~IT956-16 (Spain) and FONDOS FEDER under grant FIS2014-57956-P (Spanish government). The research of M.\,K.~was financed by the Polish National Science Center Grant DEC-2012/06/A/ST2/00395 and by a grant for the Short Term Scientific Mission (STSM) ``Multiverse impact onto the cosmic microwave background and its relation to modified gravity'' (COST-STSM-CA15117-36137) awarded by the above-mentioned COST Action. M.\,K.~and J.\,M.~would like to thank the Centro de Matem\'atica e Aplica\c{c}\~{o}es of the Universidade da Beira Interior in Covilh\~{a}, Portugal for kind hospitality while part of this work was done. M.\,K.~also thanks M.\,P.~D\k{a}browski for fruitful discussions. J.\,M.~is thankful to UPV/EHU for a PhD fellowship.

%
%

\appendix

\section{Computation of the Euclidean action integral}

%
%

\subsection{Direct derivation}
\label{App_A}

Let us consider the integral
\begin{align}
	\label{App11}
	I = \int_{x_-}^{x_+} \frac{\D x}{2x} \sqrt{(x_+ - x) (x-x_-) (x+x_0) } 
	\,.
\end{align}
To solve this integral, we employ once more the change of variable  in \eqref{Wormhole_xi}
\begin{align}
	x\rightarrow\txi = \arccos\left(\sqrt{\frac{x-x_-}{x_+-x_-}}\right).
\end{align}
After substituting in Eq.~\eqref{App11}, we obtain the following result 
\begin{align}
	\label{App12}
	I =&~ \left(x_0+x_+\right)^{3/2} \,
	\int^{\frac{\pi}{2}}_{0} {\D \txi}\,
	\frac{\Phi(\txi)}{R}
	\,,
\end{align}
where
\begin{align}
	\Phi(\txi)
	=
	\frac{
		\tilde{k}^2\kappa^2\sin^2(\txi)
		\left(1 - \sin^2(\txi)\right) 
		\left(1- \tilde{k}^2\sin^2(\txi)\right)
	}{1 - \kappa^2\sin^2(\txi)}
	\qquad
	\mathrm{and}
	\qquad
	R
	=
	\sqrt{1- \tilde{k}^ 2\sin^2(\txi)}
	\,,
\end{align}
and the parameters $\tilde{k}$ and $\kappa$ are defined as
\begin{align}
	\tilde{k}:=\sqrt{\frac{x_+ - x_-}{x_+ + x_0}}
	\qquad
	\textrm{and}
	\qquad
	\kn := \sqrt{\frac{x_+ - x_-}{x_+}}
	\,.
\end{align}
We can now expand the rational function $\Phi$ in the argument of the integral in Eq.~\eqref{App12} to write $I$ as a sum of four different integrals
\begin{align}\label{I4}
	I 
	=&~
	\left(x_0+x_+\right)^{3/2} \left[
		A_1 \int^{\frac{\pi}{2}}_{0} \frac{{\D \txi \, \sin^4(\txi)}}{R}
		+ A_2 \int^{\frac{\pi}{2}}_{0} \frac{{\D \txi \, \sin^2(\txi)}}{R}
		- A_3 \int^{\frac{\pi}{2}}_{0} \frac{\D \txi}{R}
		+ A_3 \int^{\frac{\pi}{2}}_{0} \frac{\D \txi}{\left(1-\kappa^2\sin^2(\txi)\right)R}
	\right],
\end{align}
where
\begin{align}
	A_1 = - \tilde{k}^4
	\,,
	\qquad
	A_2 = \tilde{k}^2\left[ 1  + \tilde{k}^2 \left(1 - \frac{1}{\kappa^2}\right)\right],
	\qquad
	A_3 = \tilde{k}^2\left( 1- \frac{\tilde{k}^2}{\kappa^2}\right)\left( 1- \frac{1}{\kappa^2}\right).
\end{align}
We note that the last three terms on the right hand side of Eq.~\eqref{I4} can be expressed in terms of  complete elliptic integrals of the first, second and third kind: $K(\tilde{k}^2)$, $E(\tilde{k}^2)$ and $\Pi(\kappa^2|\tilde{k}^2)$  \cite{abra,Baker1890}. This allows us to simplify Eq.~\eqref{I4} as
\begin{align}
	\label{App13}
	I
	= &~ 
	\left(x_0+x_+\right)^{3/2}\left[
	A_1 \int_0^{\pi/2} \frac{\D \txi\,\sin^4(\txi) }{R}  
	+\left(A_2- A_3\right)  K(\tilde{k}^2)
	- A_2 E(\tilde{k}^2)
	+ A_3 \Pi(\kappa^2|\tilde{k}^2)
	\right].
\end{align}
Finally, we tackle the first term on the right hand side of Eq.~\eqref{App13}. It can be proven that 
\begin{align}
	\frac{\D }{\D \txi} \left[\sin(\txi)\cos(\txi) R\right]
	= &~
	\frac{3 \tilde{k}^2 \sin^4(\txi)}{R}
	- \frac{2+ \tilde{k}^{2} }{\tilde{k}^2} \frac{1}{R}
	+ 2\,\frac{1+\tilde{k}^{2}}{\tilde{k}^2} R
	\,.
\end{align}
By integrating the previous equation in $\txi$ from $0$ to $\pi/2$, the left hand side of the equation vanishes and, after some rearrangement of terms, we can write
\begin{align}
	\label{App14}
	 \int_0^ {\pi/2} \frac{\D\txi \, \sin^4(\txi)}{R}
	=&~  
	\frac{1}{3}\frac{2 + \tilde{k}^{2}}{\tilde{k}^4} \int_0^ {\pi/2} \frac{\D\txi}{R}
	- \frac{2}{3}\frac{1+\tilde{k}^{2}}{\tilde{k}^4} \int_0^ {\pi/2}  \D\txi\,R
	\nn\\
	=&~ 
	\frac{1}{3}\frac{2 + \tilde{k}^{2}}{\tilde{k}^4} K(\tilde{k}^2)
	- \frac{2}{3}\frac{1+\tilde{k}^{2}}{\tilde{k}^4} E(\tilde{k}^2)
	\,.
\end{align}
By substituting \eqref{App13} into \eqref{App14}, we finally obtain
\begin{align}
	I
	= &~ 
	\left(x_0+x_+\right)^{3/2}
	\left[
	C_K\,  K(\tilde{k}^2)
	+ C_E\, E(\tilde{k}^2)
	+ C_\Pi\, \Pi(\kappa^2|\tilde{k}^2)
	\right],
\end{align}
where
\begin{align}
	C_K = \tilde{k}^2\left[
		\frac{1}{3}+\frac{1}{\kappa^2} 
		+ \tilde{k}^2\left(\frac{1}{3}-\frac{1}{\kappa^4}\right)
	\right],
	\qquad
	C_E = - \tilde{k}^2\left[
		\frac{1}{3}
		+ \tilde{k}^2\left(\frac{1}{3} - \frac{1}{\kappa^2} \right)
	\right],
	\qquad
	C_\Pi=A_3
	\,.
\end{align}

%
%

\subsection{Cross-check of the integral by differentiation}
\label{App_crosscheck}

In order to cross-check whether the solution to the tunnelling integral
\be
I(K) = \int_{a_-}^{a_+}\D a\,\left|\frac{1}{a} \sqrt{a^6  - \frac{a^4}{\HdS^2} + \frac{K^2}{\sigma^2 \HdS^2}}\right|
\ee
found above is correct, we differentiate the integral with respect to $K$ and use the commutativity of integration and differentiation:
\be
\frac{\partial I(K)}{\partial K} = \frac{K}{\sigma^2 \HdS^2}\int_{a_-}^{a_+}\left|\frac{\D a}{a\sqrt{a^6  - \frac{a^4}{\HdS^2} + \frac{K^2}{\sigma^2 \HdS^2}}} \right|\,.
\ee
Using the roots defined in Eqs.~(\ref{root+})--(\ref{root0}), we can write:
\be
\frac{\partial I(K)}{\partial K} = \frac{K}{\sigma^2 \HdS^2}\int_{a_-}^{a_+}\left|\frac{\D a}{a\sqrt{(a^2 - a_+^2)(a^2 - a_-^2)(a^2 + a_0^2)}}\right|\,.
\ee
We substitute $x = a^2$ as well as $x_i = a_i^2$ for $i\in\{+,-,0\}$ and obtain:
\begin{align}
\frac{\partial I(K)}{\partial K} &= \frac{K}{2 \sigma^2 \HdS^2}\int_{x_-}^{x_+}\left|\frac{\D x}{x\sqrt{(x - x_+)(x - x_-)(x + x_0)}}\right|
\nn\\
&= -\,\frac{K}{2 \sigma^2 \HdS^2}\int_{x_-}^{x_+}\frac{\D x}{x\sqrt{(x_+ - x)(x - x_-)(x + x_0)}}\,.
\end{align}
We can now apply formula (3.137:6) on p.~262 in \cite{grad} by identifying:
\be
r:=0\,,\quad \bar{a}:=x_+\,,\quad \bar{b} \equiv u:=x_-\,,\quad \bar{c}:=-x_0\,.
\ee
Note that according to p.~254 of \cite{grad}:
\be
\lambda 
:= \arcsin\left(\sqrt{\frac{\bar{a}-u}{\bar{a}-\bar{b}}}\right) 
=  \frac{\pi}{2}
\,,
\quad 
p:=\sqrt{\frac{\bar{a}-\bar{b}}{\bar{a}-\bar{c}}}
=\tilde{k}\,.
\ee
Hence, the differentiated integral reads:
\begin{align}
\frac{\partial I(K)}{\partial K} &=
-\,\frac{K}{\sigma^2 \HdS^2}\,\frac{1}{x_+\sqrt{x_+ + x_0}}\,\Pi\left(\frac{\pi}{2},
\kappa^2,
\tilde{k}\right)
\nn\\
&=-\,\frac{K}{\sigma^2 \HdS^2}\,\frac{1}{x_+\sqrt{x_+ + x_0}}\,\Pi\left({\kappa^2}\middle|\tilde{k}^2\right).
\end{align}
Plotting this function and the $K$-derivative of \eqref{tunnelint} shows that the functions are identical, their difference is zero for $0 < K < K_\text{max}$. Hence, the cross-check confirms that \eqref{tunnelint} is the correct expression for the integral.

%
%

\bibliographystyle{apsrev}

\end{document}